# Mixtures of 1-butyl-3-methylimidazolium tetrafluoroborate ionic liquid and acetonitrile: a molecular simulation

**Vitaly V. Chaban and Oleg V. Prezhdo**

Department of Chemistry, University of Rochester, Rochester, New York 14627, United States.
E-mail: v.chaban@rochester.edu

Recently, we introduced a new force field (FF) to simulate transport properties of imidazolium-based room-temperature ionic liquids (RTILs) using a solid physical background. In the present work, we apply this FF to derive thermodynamic, structure, and transport properties of the mixtures of 1-butyl-3-methylimidazolium tetrafluoroborate, [BMIM][$BF_4$], and acetonitrile (ACN) over the whole composition range. Three approaches to derive a force field are formulated based on different treatments of the ion-ion and ion-molecule Coulomb interactions using unit-charge, scaled-charge and floating-charge approaches. The simulation results are justified with the help of experimental data on specific density and shear viscosity for these mixtures. We find that a phenomenological account (particularly, simple scaled-charge model) of electronic polarization leads to the best-performing model. Remarkably, its validity does not depend on the molar fraction of [BMIM][$BF_4$] in the mixture. The derived FF is so far the first molecular model which is able to simulate all transport properties of the mixtures, comprising RTIL and ACN, fully realistically.

## Introduction

Room-temperature ionic liquids (RTILs) are a broad class of salts, which are formed by large asymmetric organic cation and organic or inorganic anion[1-9]. The total number of RTILs, synthesized up to date, exceeds 500, and still vigorously increases. Due to their remarkable diversity, ionic liquids possess a variety of robust applications[1,2,9], the most known of which are electrolytes for electrochemical devices[7,10], reaction medias[11,12], lubricants[13], and solvents for synthetic and catalytic schemes[14,15].

Most contemporary electrolytes exploit ethylencarbonate, dimethylcarbonate and diethylcarbonate as solvents, along with lithium hexafluorophosphate as a source of ions[16]. Although such performant systems permit a large amount of charging/discharging cycles without significant loss in capacity, their volatility and flammability present a serious safety problem. As a result, the above electrolytes can not be applied, if the working temperature of the device exceeds 330 K. Owing to their wide electrochemical windows (> 4 V)[17], wide liquid state temperature ranges, negligible vapor pressure and, importantly, non-flammability, RTILs are considered a budding alternative to traditional non-aqueous electrolytes. However, neat RTILs are highly viscous (more than 30 cP at 300 K) substances[18-21]. Moreover, they exhibit quite low ionic conductivity, which implies an extremely hindered mobility of the particles. In order to overcome this con, organic dipolar aprotic solvents can be added to ionic liquids. On a related note, RTILs are able to dissolve polymers with a formation of ion conducting composites[22,23] for electrochemical applications[7,17,19,24,25].

In the present work, we consider the mixtures (with compositions between 0-100%) of 1-butyl-3-methylimidazolium tetrafluoroborate ([BMIM][$BF_4$]) and acetonitrile (ACN) (Figure 1) using classical molecular dynamics (MD) simulations. The investigation is carried out by means of exclusively conventional, non-polarizable force field models. Three approaches for constructing the best-performing force field (FF) for such kind of binary systems are considered. The available experimental densities[26] and viscosities[26] of the mixtures of [BMIM][$BF_4$] and ACN at 298 K are applied to validate the resulting models.

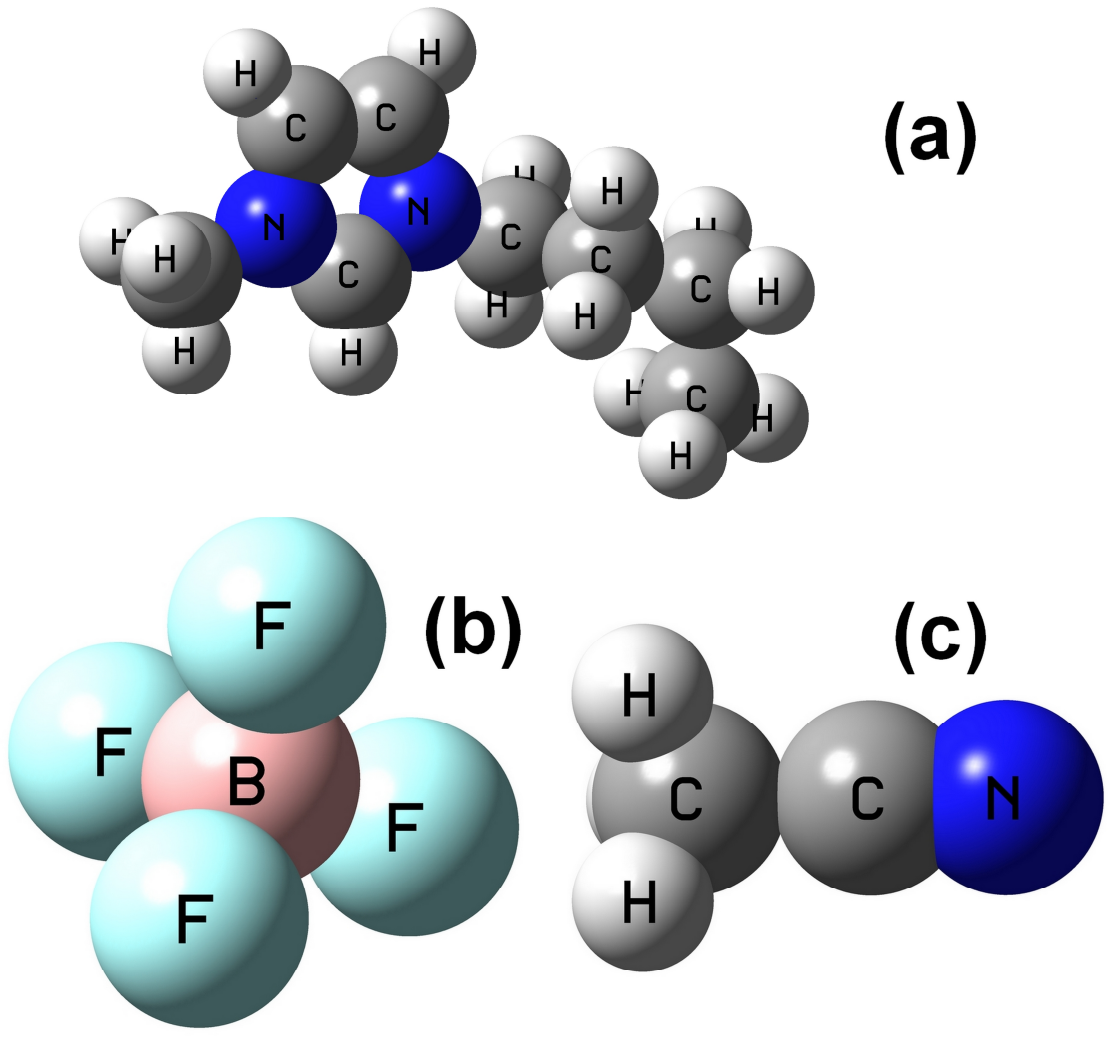

Figure 1. The simulated particles, a) 1-butyl-3-methylimidazolium$^+$, b) tetrafluoroborate$^-$, c) acetonitrile.

We found that the model providing a phenomenological account for an electronic polarization in ionic liquid performed the best[27]. Importantly, its applicability is independent on the composition of the mixture. The self-diffusivities of all components, ionic conductivities, shear viscosities, specific densities, heats of vaporization, and structure pair correlation functions are reported. Based on the simulation results, it is demonstrated that mixtures of [BMIM][$BF_4$] and ACN with a small content of RTIL allow to increase conductivity by approximately a factor of ten, and consequently, to decrease viscosity by about a factor of 40 (at 298 K), as compared with neat RTILs. An ability to significantly enlarge ionic conductivity by varying exclusively the molar fractions of the mixture components favors the applications of [BMIM][$BF_4$] + ACN systems as novel non-aqueous electrolytes[7].

## Methodology

The phase trajectories for eight systems (Table 1), containing 0, 5, 10, 25, 50, 75, 90, and 100 molar percents of [BMIM][$BF_4$], are derived by means of classical MD simulations with pairwise interaction potentials. The molecular dynamics runs are accomplished using the GROMACS 4.0 simulation package[28] in the constant temperature and constant pressure ensemble. All atoms of 1-butyl-3-methylimidazolium tetrafluoroborate and acetonitrile, including hydrogens, are represented as separate interaction centers, possessing Lennard-Jones (12,6) parameters and electrostatic charges. The constant temperature of 298 K is maintained using the V-rescale thermostat[29] with a response time of 1.0 ps. The constant pressure (1 bar) is maintained using the Parrinello-Rahman technique for pressure coupling[30] with a relaxation time of 4.0 ps. The widely used leap-frog algorithm is used to integrate the equations of motion with a time-step of 1 fs.

Table 1. System sizes of [BMIM][$BF_4$] ($N_1$) and ACN ($N_2$), shear viscosities and ionic conductivities at 298 K and 1 bar derived using three models (M.1, M.2, and M.3) of the simulated mixtures. The small indices correspond to the standard deviations of the results

| System size | | | Shear viscosity, cP | | | Ionic conductivity, S/m | | |
|---|---|---|---|---|---|---|---|---|
| $N_1$ | $N_2$ | $x_1$, % | M.1 | M.2 | M.3 | M.1 | M.2 | M.3 |
| 0 | 300 | 0 | $0.4_1$ | $0.4_1$ | $0.4_1$ | 0.0 | 0.0 | 0.0 |
| 15 | 285 | 5 | $0.8_1$ | $0.6_1$ | $0.8_2$ | $2.0_2$ | $2.4_2$ | $1.9_2$ |
| 30 | 270 | 10 | $1.1_2$ | $1.0_2$ | $1.1_2$ | $1.7_2$ | $3.0_3$ | $2.4_4$ |
| 75 | 225 | 25 | $8_3$ | $2.3_6$ | $4.2_6$ | $1.1_1$ | $3.2_2$ | $1.25_8$ |
| 150 | 150 | 50 | $71_{15}$ | $12_3$ | $28_5$ | $0.25_1$ | $1.4_1$ | $0.60_8$ |
| 225 | 75 | 75 | ~400 | $43_5$ | $46_4$ | $0.06_1$ | $0.53_6$ | $0.33_4$ |
| 270 | 30 | 90 | ~800 | $63_9$ | $71_9$ | $0.04_1$ | $0.35_4$ | $0.28_2$ |
| 300 | 0 | 100 | ~1300 | $98_{11}$ | $98_{11}$ | $0.04_1$ | $0.31_2$ | $0.31_2$ |

The slow dynamics of particles in RTILs requires longer relaxation and production stages than it is usually performed in the classical MD studies. It is also necessary for the mixtures with small molar fractions of the ionic liquid, since less amount of particles (ions) requires longer simulation times to achieve ergodicity. In the present study, the initial relaxation is carried out at 350 K (for molar fractions of RTIL more than 50%) and 320 K (for molar fractions of RTIL less and equal to 50%) during 4,000 ps in the constant volume constant temperature ensemble. Next, the obtained systems are gradually cooled to the target temperature (298 K) during 2,000 ps. The transport properties under discussion are derived from 25,000 ps trajectories for each system. A few consequent MD trajectories of 25,000 ps are used to obtain reliable averages for ionic conductivity and shear viscosity values. To ensure extensive trajectory sampling, each productive run begins with an assignment of random velocities to ions, *i.e.* the generated trajectories are statistically independent.

The simulated systems of mixtures and pure components ([BMIM][$BF_4$] and ACN) (Table 1) are placed in cubic MD boxes with periodic boundary conditions applied along all three Cartesian directions. The long-range electrostatic forces are treated by the Particle Mesh Ewald method[31] with the cut-off distance for real-space component equals to 1.5 nm. The Lennard-Jones part is treated by the conventional shifted force method with a switch region between 1.2 nm and 1.3 nm. The list of neighbors during the simulation is updated every 0.01 ps (10 time-steps) within a sphere of radius of 1.5 nm.

### Force fields

A number of force fields for RTILs[20,24,32-42], including [BMIM][$BF_4$], have been suggested during the last decade covering a broad spectrum of species. Both empirical FFs and direct *ab initio* calculations of the ion pairs, were extensively used to derive structure, dynamics, and energetic properties[18,20,21,24,43-49] of pure RTILs and certain mixtures of practical interest. Borodin suggested a polarizable force field[24] employing the Thole-type approach. This model overcomes most of the problems of the conventional non-polarizable force fields, but instead, it is significantly more computationally expensive. Recently, it was numerically proven[27] that ionic dynamics of imidazolium-based RTILs, in order to be reproduced realistically, requires a uniform decrease of electrostatic charges on both ions. In turn, the initial charges can be either taken from the systematically derived FFs or calculated independently using R(ESP) methodology, which reflects the electrostatic potential around a particle usually based on the high level *ab initio* calculations.

However, binary mixtures of molecular and ionic solvents are significantly more complex systems than pure RTILs. Owing to a new kind of interactions, *i.e.* cation-acetonitrile and anion-acetonitrile, varying affinity of the cations and anions to the solvent neutral molecules and varying local ion-molecular structure at different mixture compositions, such simulation is still a challenge for computational society. So far, just a single computational study of the [BMIM][$BF_4$] + ACN mixtures has been published[49]. The authors constructed their own models both for acetonitrile and RTIL and carried out classical equilibrium MD simulations over the whole range of compositions. Although a reasonable agreement with experimental viscosities[26] is reported, the simulated transport constants are significantly overestimated. The productive stage, persisting for only 200 ps, is definitely not enough to overpass the sub-diffusion trajectory part for mean square displacements[27] in the case of imidazolium-based ionic liquid. As a result, the reported values are realistic notwithstanding the unphysical simulation methodolody. Nevertheless, the model from Ref. [49] cannot be applied for reliable simulations of [BMIM][$BF_4$] + ACN mixtures. In the present investigation, we consider a set of force field models, which represent different assumptions concerning the Coulombic interactions in these binary mixtures.

First, the unit-charge FF for [BMIM][$BF_4$] from Ref.[36,49] with integral electrostatic charges on cation and anion (model 1) is applied. The charges on all interaction sites of the ions have been derived by separate fitting electrostatic potentials for these particles in vacuum. Thus, the sum of partial electrostatic charges on $BMIM^+$ and $BF_4^-$ are +1 and -1, respectively.

Second, the FF[27] for [BMIM][$BF_4$] with uniformly scaled charges (model 2) is considered. The periodic density functional theory computations of electrostatic potential have been used to derive the scaling factor, $f_{sc}$, of 0.81 for this RTIL as argued in Ref. [27]. Thus, the sum of partial electrostatic charges on $BMIM^+$ and $BF_4^-$ are +0.81 and -0.81, respectively. Note, that model 2 is a refined version of model 1.

Third, whereas model 1 does not consider polarization of electronic shells at all, model 2 provides a phenomenological account for this phenomenon in bulk RTIL. According to intuition, in condensed phase, the polarization mainly occurs owing to neighboring counterions, which tend to communalize their valence electrons. Additional component (ACN), which is present in the simulated mixture, significantly complicates the situation, especially if the whole composition range of the binary mixture is considered. Indeed, while the molar fractions of [BMIM][$BF_4$] are small, the ions are surrounded primarily by neutral molecules. As the molar fraction of RTIL increases, the amount of counterions in the first coordination shell of the ions increases accordingly. Ultimately, this should lead to a varying degree of electronic polarization of [BMIM][$BF_4$] for different compositions of the mixture. In order to deal with this challenge within the framework of non-polarizable force field for classical molecular dynamics simulation, we introduce a floating scaling factor. In this approach, the value of $f_{sc}$ is a function of the molar composition of the simulated mixture (model 3).

In model 3, it is *a priori* assumed that counterions are polarized exclusively by one another. Hence, for infinitively small molar fraction of RTIL, $f_{sc} = 1$. In turn, for infinitively small molar fraction of ACN, $f_{sc}=0.81$[27]. For intermediate compositions, the scaling factor is introduced as follows, $f_{sc}^{mix}=\sqrt{(1-f_{sc}^2)(1-x)+f_{sc}^2}$, where $f_{sc}$ is a scaling factor for the pure RTIL, and $x$ is a molar fraction of the RTIL in the mixture. The present functional form arises from the linear dependence of the ion-ionic and ion-molecular interaction energies (e.g. heat of vaporization discussed below) on the molar fraction. This model considers uniform distribution of the ions throughout the mixture, and does not account for any possible ionic cluster formation.

It should be noted that the recent FF model[27] for RTIL provides a slightly lower density due to the decrease of the inter-ionic interaction energies. Generally, the density vigorously depends on the van der Waals interactions, *i.e.* the parameters in the Lennard-Jones equation. An additional adjustment of density of the imidazolium-based RTILs has been performed by Liu *et al.*[36] by comparing parameters from different force fields and conducting *ab initio* calculations of the ion pairs. According to that work, significantly different parameters for the hydrogen atoms (HA/H4/H5) of imidazole ring are suggested by AMBER ($r_{min}$=2.432 nm), CHARMM ($r_{min}$=1.247 nm), and OPLS ($r_{min}$=2.420 nm) force fields. It is reported that the best density of liquid [BMIM][$PF_6$] is provided by the AMBER FF (1331 kg/m$^3$). Next, the geometry optimization of 1,3-dimethylimidazolium hexafluorophosphate has been performed at HF/6-31+G(d) level. The *ab initio* results suggest an asymmetric geometry of the ion pair, that molecular mechanics (MM) fails to reproduce. As a result, the length of hydrogen bond between the cation and the anion is significantly larger in MM, if $\sigma$=2.511 Å is applied. It is shown[36] that the asymmetry can be reproduced using MM, while $r_{min}$ (H5) is as small as 1.0 Å. A hydrogen bond, C-H...F, with $r$(C-H)=2.17 Å is obtained. The better fit of other geometrical parameters is observed as well. Finally, $\sigma$(H5)=1.782 Å is assumed for imidazolium-based cations. Meanwhile, it is not clear why this refinement for H5 is not introduced for H4, although these sites are nearly chemically identical. For instance, in the OPLS FF both H4 and H5 are simulated with the same atom type, HA. The decrease of the diameters of the imidazole ring hydrogen atoms leads a noticeable increase of the ionic liquid density[36].

Our simulation of density of [BMIM][$BF_4$], using $f_{sc}=0.81$, provides 1164, 1169, and 1160 kg/m$^3$ at 298 K for AMBER, CHARMM, and OPLS, respectively. Employing $f_{sc}=1.00$, we obtain 1205, 1206, and 1202 kg/m$^3$ for the same force fields. Note, the experimental value according to Wang *et al.*[26] is 1211 kg/m$^3$, *i.e.* all FFs and scaling factors provide systematically underestimated values. Unlike the Liu's case with [BMIM][$PF_6$], CHARMM operates best for [BMIM][$BF_4$] (1206 kg/m$^3$). This conclusion is true both for $f_{sc}=1.00$ and $f_{sc}=0.81$. Thus, varying the parameters for the HA interaction sites, it is principally possible to tune the simulated density for the scaled-charge models. Nevertheless, we do not apply this refinement in the productive simulations, since its physical background is not evident.

The last published six-site model of Nikitin[50] is used to simulate acetonitrile. Likewise, it is based on the parameters of the AMBER FF. The model excellently reproduces density (773 kg/m$^3$), heat of vaporization (33.5 kJ/mol) structural distributions[50], and shear viscosity (~0.4 cP) of bulk ACN, as well as the corresponding properties of the acetonitrile-water mixtures. However, diffusion constant at 298 K appears underestimated ($3.4\times10^{-9}$ m$^2$/s) as compared to the experimental value ($4.3\times10^{-9}$ m$^2$/s). This con should be taken into account when the transport properties of the RTIL+ACN mixtures are considered.

Since both the model of RTIL and the model for ACN utilize AMBER force field, the appropriate combination rules for Lennard-Jones interactions and scaling 1-4 interactions are prescribed. Note, the three above models differ only by the representation of electrostatics and use the same Lennard-Jones (12,6) and bonded parameters as the previous works suggest[36,50]. Certainly, the interaction energy decrease may alter the intra-ionic oscillations of the bonded atoms. Here, we assume that this impact is negligible and do not incorporate it into the models 2 and 3.

**Calculated properties**

We derive distribution functions, heat of vaporization, and transport properties (diffusion constant, shear viscosity and ionic conductivity), writing down the atomic coordinates and interaction energies every 0.02 ps (20 time-steps). Diffusion constants and their standard deviations are found using the consequent parts of trajectory of 5,000 ps. In turn, shear viscosity and ionic conductivity require longer times (25,000 ps) to obtain reliable values. The convergence of all transport properties is assured by calculating them as a function of simulation time (Figure 2). A few consequent runs with randomly generated velocities, as mentioned above, are applied to estimate the averages and standard deviations of the resulting values. In turn, structure pair correlation functions, specific density and heat of vaporization can be derived using significantly poorer sampling (1,000 ps), provided that the simulated systems are equilibrated thoroughly.

Density of the systems of RTIL (Figure 3) and its fluctuation during the simulation are estimated from the oscillations of the MD box volume. Shear viscosity, $\eta$, (Figure 4, Table 1) is obtained by integrating the autocorrelation function of the off-diagonal elements of pressure tensor. This method allows for the viscosity of the system to be obtained using the equilibrium MD methodology, although extensive sampling is required. Diffusion constant (Figure 5, Table 2), $D$, is computed *via* the Einstein relation, through plotting mean square displacements of all atoms of each species. Ionic conductivity, $\sigma$, (Figure 6, Table 2) is obtained in the framework of the Einstein-Helfand formalism from the linear slope of mean-square displacements of the collective translational dipole moment. Importantly, for this method to work correctly, periodic boundary conditions should be removed prior to computation, *i.e.* "free" diffusion is necessary. The corresponding formulas for the transport properties can be found in Ref. [27]. Heat of vaporization (Figure 7), $H_{vap}$, is estimated at 298 K only. Importantly, vapor phase of RTILs is assumed to contain only neutral ionic pairs, [BMIM][$BF_4$], which do not interact with one another. Radial distribution functions (Figures 8-9), $g_{ij}(r)$, are calculated using a classical definition on the trajectory parts of 1,000 ps.

**Results and Discussion**

Although RTILs exist in a liquid state at room temperature, they differ from the conventional liquids due to their lack of particle (ionic) dynamics. Consequently, molecular dynamics simulations of RTILs require longer relaxation and production stages in order to provide a proper sampling of the phase trajectory. On a related note, special attention should be given to avoid the calculation of dynamical properties from the sub-diffusion region of mean square displacements. Figure 2 depicts the diffusion constant of $BMIM^+$, shear viscosity and ionic conductivity of the mixture, containing 90% of [BMIM][$BF_4$] and 10% of ACN, *i.e.* the mixture with presumably the slowest ionic and molecular motions. Clearly, the diffusion constant ($D$), calculated using the parts of trajectory shorter than 1,000 ps, is a few times larger than the $D$ calculated using longer trajectories. Importantly, this behavior does not noticeably depend on $f_{sc}$ (1.00 or 0.81), although the particular values differ significantly. For the trajectories parts, which are longer than 2,000 ps, $D$ exhibits a converged value. Insignificant (< 20%) fluctuations around 0.20 and 2.5 ($\times 10^{-9}$ $m^2/s$) are observed as the simulation time increases. In turn, $\eta$ and $\sigma$, being the collective properties of the system, require more extensive sampling (10,000 ps and more). Whereas the shear viscosity, derived *via* the Einstein formula, gradually increases over time, the deviations of the ionic conductivity are random. The dependences of both $\eta$ and $\sigma$ over time do not exhibit a clear dependence on $f_{sc}$. Meanwhile, the conductivities provided by model 1 are *ca.* 10 times lower than those by model 2 (0.04±0.01 *vs.* 0.35±0.04 S/m). The situation for viscosity is inverse (~800 *vs.* 63 cP). Based on such an analysis of the mixture, containing 90% of [BMIM][$BF_4$] and 10% of ACN, we select the production stages of 25,000 ps for all considered systems. Since other MD systems contain a lower amount of RTIL, and consequently, should exhibit considerably faster dynamics, the chosen simulation time is enough for sure.

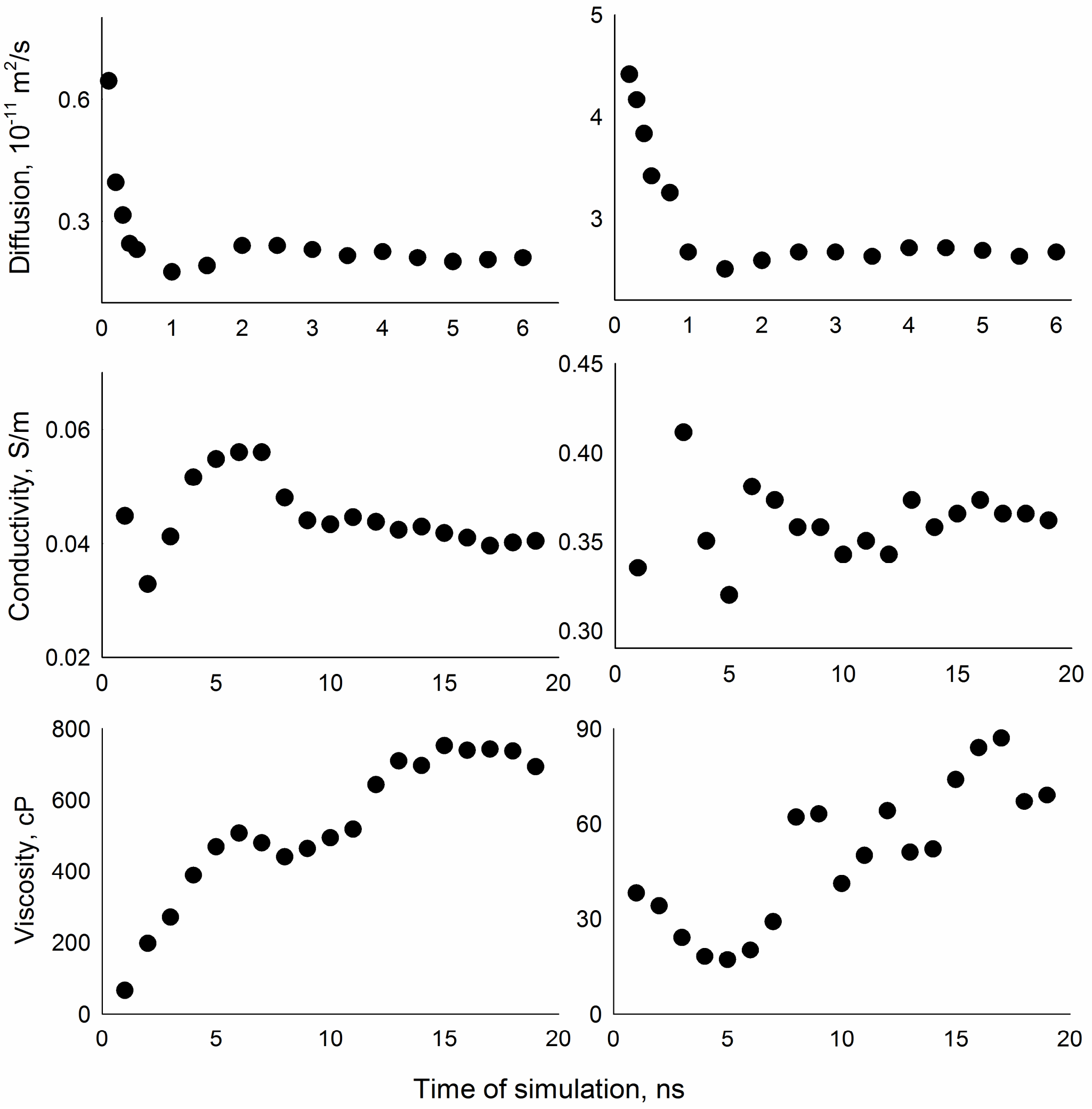

Figure 2. Transport properties of the mixture containing 90% of [BMIM][$BF_4$] and 10% ACN as a function of simulation time computed at 298 K and 1 bar using model 1 (left) and model 2 (right).

The specific density of the mixture (Figure 3) exhibits a significant deviation from the ideal solution. The shapes of the lines suggest the negative non-ideality in the whole range of compositions. In turn, this indicates that mixing of ACN and [BMIM][$BF_4$] leads to a more efficient packing and stronger attractive interparticle interactions than in the pure compounds. All force field models capture this behavior well, athough numeric values slightly differ (Figure 3). As shown in Ref. [27], the decrease of $f_{sc}$ leads to the decrease of density by 1-3% for neat RTIL. The experimental densities are reported in Ref.[26] and are the linear functions of the mass fraction of RTIL in the mixture. Note, the molar fraction, $x$, is connected to mass fraction, $\omega$, as $x = M^s \times \omega$, where $M^s$ is a molecular mass of the mixture. Overall, the simulated densities are of satisfactory accuracy over the whole range of compositions regardless of the force field used. Whereas the underestimation of density by models 2 and 3 is 2-3%, the deviations for model 1 (non-scaled charges) are within 1.5 %.

The reported results for model 1 are slightly different from Ref. [49] owing to the different model of ACN. It should be underlined that the new model of Nikitin provides a better density of the neat acetonitrile (773 vs. 748 kg/m$^3$) due to fine parametrization. The underestimation of density for the mixtures with larger molar fractions of RTIL come directly from the similar underestimation for the neat [BMIM][$BF_4$] liquid. On a practical note, the negative deviation from the ideal behavior (Figure 3) presumes that the overall mobility of particles in the mixture should drastically increase and viscosity should accordingly decrease when the molar fraction of RTIL is 25% or less.

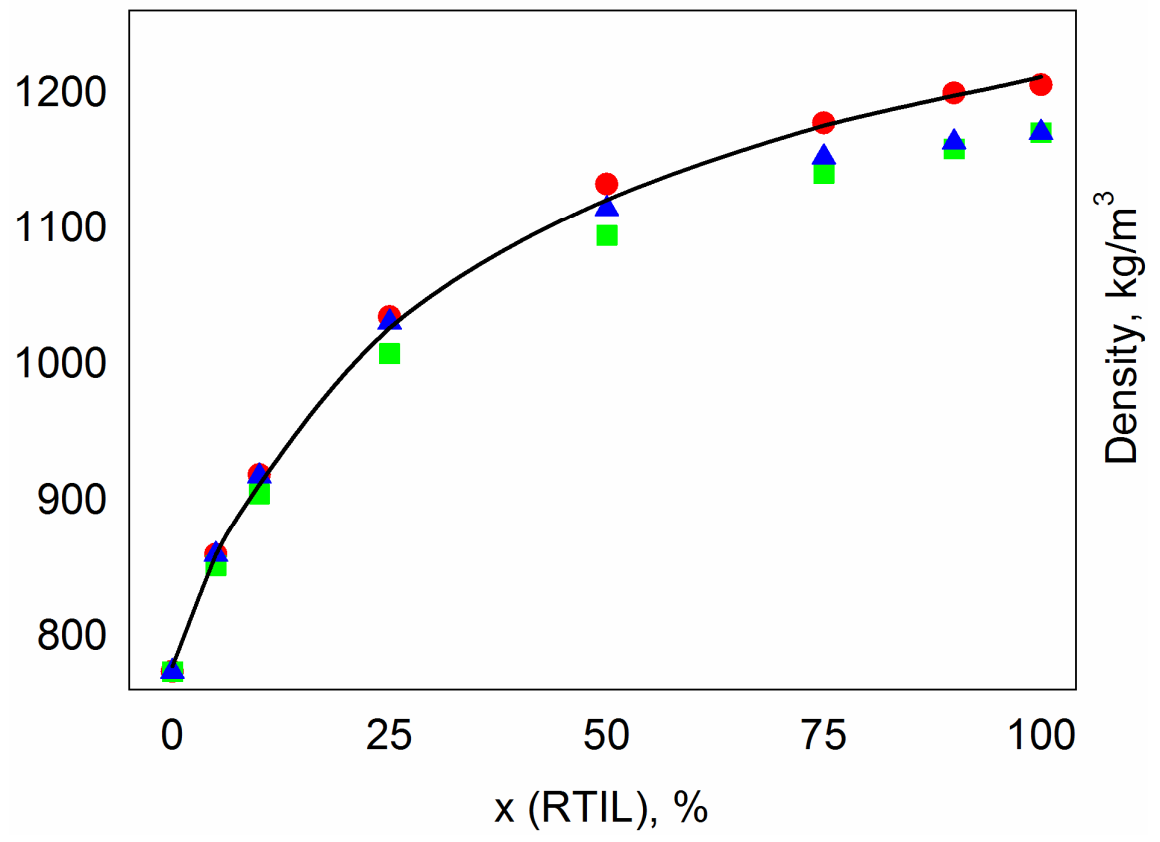

Figure 3. Specific density of the [BMIM][$BF_4$] + ACN mixtures computed at 298 K and 1 bar for model 1 (red circles), model 2 (green squares), and model 3 (blue triangles). The experimental values are presented by solid black line.

Shear viscosity, $\eta$, (Figure 4) of the mixtures of RTILs with a molecular solvent can be easily measured experimentally. Hence, it presents a convenient tool to parametrize the new FFs for these systems. While most properties, used to parametrize FFs, are thermodynamic, viscosity is especially valuable since it represents transport properties of the systems. As we show below, an ability of certain force field to simulate density and thermodynamics reliably does not guarantee the realism of the microscopic motion. Unfortunately, in the previous simulations Wu *et al.*[49] did not derive the viscosities for [BMIM][$BF_4$]+ACN systems directly, applying the Stokes-Einstein equation instead. The results, obtained in the present work for all three models of mixtures (Figure 4), show that model 1 overestimates viscosity very significantly. Interestingly, this is true even for the smallest molar fraction (5%) of [BMIM][$BF_4$]. The calculated value is 0.8 cP, while Wang *et al.*[26] report an experimental $\eta$=0.6 cP. This viscosity is almost two times larger than of pure ACN (0.34 cP). Importantly, it suggests that even small admixtures of RTIL considerably influence the properties of the low-viscous aprotic solvent. Indeed, when acetonitrile is mixed with RTIL, strong ion-dipole interactions occur among the solute molecules and the imidazole ring of the cation. To some extent, they reduce the hydrogen bonding between the cation and the anion in [BMIM][$BF_4$] leading to the negative excess molar volumes, higher ionic mobility and decreased shear viscosities. This fact argues that the ions of RTIL are well dispersed throughout the mixture and presumably form stable solvates with the acetonitrile molecules. This supposition is not however unambiguously supported by the microscopic structure analysis, as shown below (Figures 8-9). Although model 1 performs very well in predicting densities, unfortunately it fails to reproduce experimental viscosity (Figure 4). Starting from 25% of RTIL, the viscosities of the mixtures are two and more times larger than the experimental η. One can conclude that the electrostatic potential, calculated in vacuum, definitely overestimates the corresponding energy in condensed phase, resulting in excessive non-bonded interactions[27].

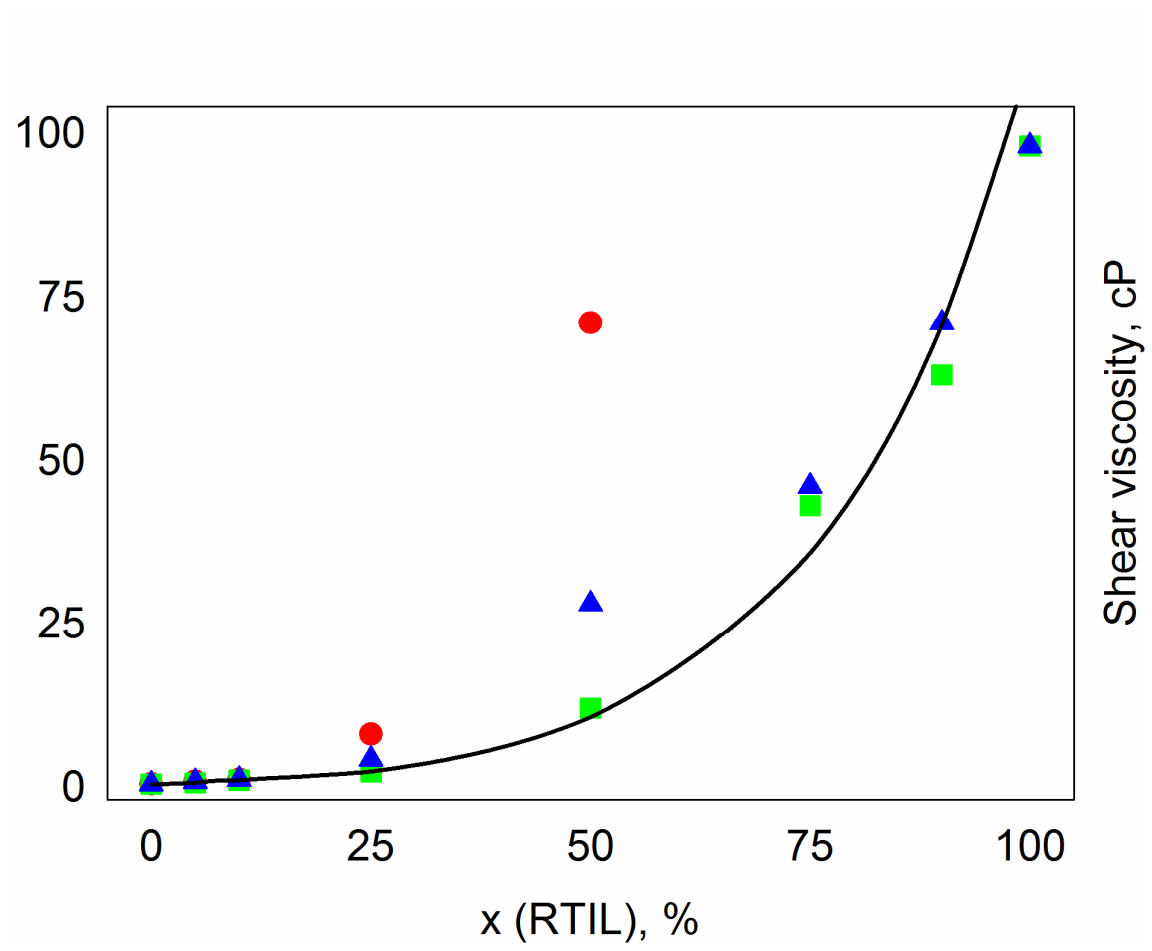

Figure 4. Shear viscosity of the [BMIM][$BF_4$] + ACN mixtures computed at 298 K and 1 bar for model 1 (red circles), model 2 (green squares), and model 3 (blue triangles). In the case of model 1, the values for 75, 90 and 100% of RTIL are so large that they do not fit into the plot (see Table 1 for these values). The experimental values are represented by means of solid black line.

Unlike model 1, models 2 and 3 perform fairly well over the whole range of compositions. Whereas model 2 performs better for 50% RTIL + 50% ACN, an inverse situation is observed for 90% RTIL + 10% ACN. Probably this scattering of the results comes from the expected fluctuations of the components of pressure tensor in the relatively small systems (a few thousand of particles), and only to a lesser extent from the particular models. In the case of the molar fractions not exceeding 50%, a systematically better performance should be granted to model 2 rather than to model 3. Surprizingly, the floating-charge model of the mixture shows less realistic values than the scaled-charge model. As a result, an assumption that ions are polarized mainly by one another and are not noticeably polarized by the solvent molecules (model 3) is incorrect. Based on the obtained results, one may conclude that the ions of RTILs are nearly equally polarized by ACN as well as by each other. In order to find a direct support of the speculation, we perform *ab initio* calculations of the electrostatic potential for the solvate complexes of $BMIM^+$ and $BF_4^-$ in ACN. The derived electrostatic potential is approximated by the partial charges on each atom including hydrogens. The calculations are accomplished on the MP2 level of theory using the 6-311++G** basis set. Remarkably, the total charges on $BMIM^+$ are 0.975, 0.967, 0.990, 0.969, 0.921, 0.963, 0.939, and 0.941, provided that the solvate contains one ions surrounded by 1, 2, 3, 4, 5, 6, 7, and 8 acetonitrile molecules, respectively. The total charges on $BF_4^-$ are 0.972, 0.939, 0.906, 0.858, 0.837, and 0.807, provided that the solvate contains one ions surrounded by 1, 2, 3, 4, 5, and 6 acetonitrile molecules, respectively. Evidently, the high-level first principles study predicts a significant charge transfer for both ions. At the same time, the amount of the tranferred charge strongly depends on the polarity of the ions. These results support our indirect findings that the electronic polarization due to solvent is nearly comparable with the polarization due to counterions. The additional investigations of the ion pairs in the solvent may be important to provide a more detailed description of this complicate phenomenon.

Based on the analysis of density and shear viscosity, the best-performing model is model 2 which uses the non-unit charges on the counterions. The model 3 also performs well and is even better for density than model 2 at a low content of RTIL. However, for the same compositions it provides a noticeably overestimated viscosity. Although model 1 reproduces the experimental density better than models 2 and 3, it is completely uncompetitive in the case of viscosity. Hence, model 1 cannot be used to analyze the other properties which are based on the ionic dynamics, *i.e.* diffusion and conductivity. Below, the results for all three force field models are listed, but primary attention is granted to model 2.

The average self-diffusion coefficient of the system, $D_{av}$, is inversely proportional to its shear viscosity, as postulated by the Einstein-Stokes relationship. Figure 5 and Table 2 demonstrate diffusion coefficients for $BMIM^+$, $BF_4^-$, ACN molecules, and $D_{av}$.

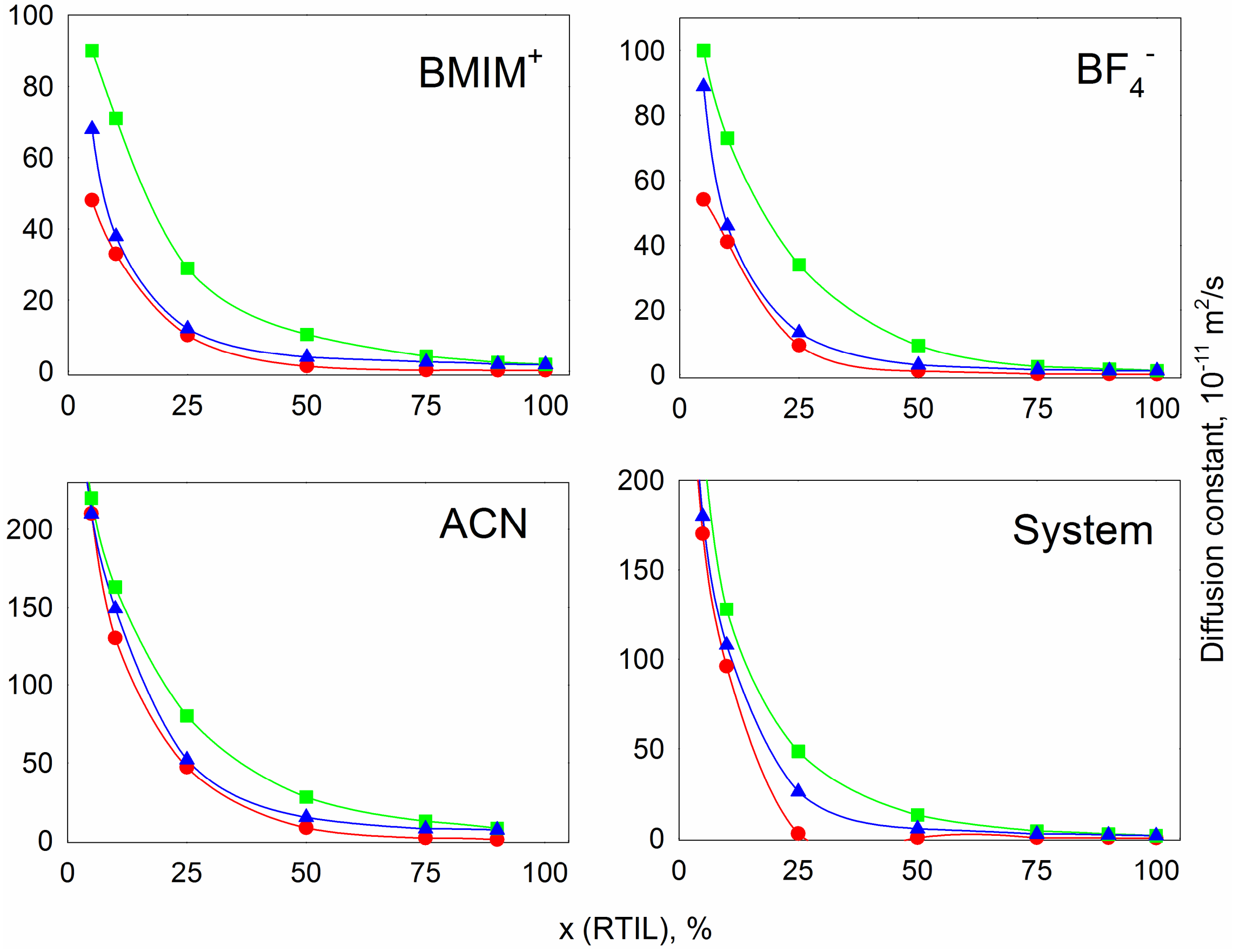

Figure 5. Diffusion constants of cation ($BMIM^+$), anion ($BF_4^-$), solvent molecules (ACN) and average diffusion constants of the [BMIM][$BF_4$] + ACN mixtures computed at 298 K and 1 bar for model 1 (red circles), model 2 (green squares), and model 3 (blue triangles). The numeric values are summarized in Table 2.

All compositional dependences can be well describes by the exponential decay analytical functions with three parameters, $D = D_0 + A \times \exp(-B \times D)$, where $D_0$ is a diffusion coefficient of pure component, and $A$ and $B$ are empirical constants. The correlation coefficient in all cases exceeds 99.5%, indicating a high accuracy of the calculated data. Remarkably, the ratio between the cationic and anionic $D$ alters as the molar fraction of [BMIM][$BF_4$] increases. If $x$(RTIL)=5%, $D$($BMIM^+$)=0.9 and $D$($BF_4^-$)=1.0 ($\times 10^{-9}$ $m^2/s$), *i.e.* the anion moves faster than the cation. Although the difference is comparable with one standard deviation, the trend is undoubted. In turn, if $x$(RTIL)=90%, $D$($BMIM^+$)=2.5 and $D$($BF_4^-$)=1.7 ($\times 10^{-11}$ $m^2/s$), *i.e.* the ratio is inverse. The molar fraction of RTIL at which the anionic transport becomes faster than the cationic transport is 25%. This observable fact can be understood as follows. If the content of ACN is small, $BMIM^+$ and $BF_4^-$ form large ionic clusters, whose structure is determined by the strong[36], although somewhat controversial[51], interactions, between fluorine atoms of the anion and hydrogen atoms belonging to the aromatic ring of the cation. In these systems, like in the pure [BMIM][$BF_4$], the self-diffusion of $BMIM^+$ is slighly bigger than $D$ of anion. As the content of ACN increases, the solute molecules tend to substitute anions in the cationic solvation shells. Therefore, a part of anions is released from the ionic cluster, and this leads to their faster motion in the mixture.

Since the anions are lighter and more compact than the cations, their motion is more significantly accelerated upon dilution. In turn, $D$(ACN) in the mixtures significantly decreases as compared to the bulk solvent. For instance, even at $x$(RTIL)=5%, $D$(ACN)=$2.2\times10^{-9}$ $m^2/s$ (model 2) that is one and a half times smaller than in bulk ($3.4\times10^{-9}$ $m^2/s$). These observations (Table 2) are practically important since they predict the range of the compositions of the mixture where ionic motion is controlled by the aprotic solvent, containing mobile molecules, rather than by RTIL containing inert ions. It is expected that diffusion coefficients of all particles and at all compositions increase from model 1 to model 3 and then to model 2. While the difference between model 1 and models 2 and 3 is significant at large parts of RTIL, it greatly decreases as the part of [BMIM][$BF_4$] decreases.

Table 2. The self-diffusion coefficients (in $\times10^{-11}$ $m^2/s$) of $BMIM^+$ ($D_+$), $BF_4^-$ ($D_-$), ACN ($D_0$), and the system as a whole ($D_{av.}$) at 298 K and 1 bar derived using three models (M.1, M.2, and M.3) of the simulated mixtures. The small indices correspond to the standard deviations of the results

| $x_1$, % | M.1 | | | | M.2 | | | | M.3 | | | |
|---|---|---|---|---|---|---|---|---|---|---|---|---|
| | $D_+$ | $D_-$ | $D_0$ | $D_{av.}$ | $D_+$ | $D_-$ | $D_0$ | $D_{av.}$ | $D_+$ | $D_-$ | $D_0$ | $D_{av.}$ |
| 0 | — | — | $340_{15}$ | $340_{15}$ | — | — | $340_{15}$ | $340_{15}$ | — | — | $340_{15}$ | $340_{15}$ |
| 5 | $48_5$ | $54_4$ | $210_{10}$ | $170_{10}$ | $90_5$ | $100_{10}$ | $220_{10}$ | $220_{10}$ | $68_7$ | $89_{10}$ | $210_{10}$ | $180_{10}$ |
| 10 | $33_2$ | $41_2$ | $130_{10}$ | $96_6$ | $71_6$ | $73_4$ | $163_3$ | $128_5$ | $38_5$ | $46_6$ | $149_5$ | $108_4$ |
| 25 | $10.1_5$ | $9_1$ | $47_3$ | $23_1$ | $29_2$ | $34_1$ | $80_{10}$ | $49_2$ | $12_1$ | $13_1$ | $52_2$ | $26_1$ |
| 50 | $1.4_2$ | $1.1_1$ | $8.2_2$ | $2.5_1$ | $10.4_5$ | $9.0_3$ | $28_3$ | $13_1$ | $4.0_2$ | $3.0_2$ | $15_2$ | $5.4_3$ |
| 75 | $0.24_2$ | $0.15_1$ | $1.7_2$ | $0.26_2$ | $4.2_1$ | $2.5_1$ | $12.5_3$ | $4.1_2$ | $2.6_1$ | $1.5_1$ | $7.6_5$ | $2.4_2$ |
| 90 | $0.20_1$ | $0.12_1$ | $0.7_2$ | $0.23_1$ | $2.5_2$ | $1.7_2$ | $8_1$ | $2.4_3$ | $2.0_3$ | $1.2_2$ | $7_1$ | $1.9_3$ |
| 100 | $0.19_3$ | $0.08_1$ | — | $0.13_1$ | $1.9_1$ | $1.2_1$ | — | $1.5_1$ | $1.9_1$ | $1.2_1$ | — | $1.5_1$ |

The ionic conductivity, σ, (Figure 6) is of extreme importance for the electrolyte applications of RTILs. For a practical use of [BMIM][$BF_4$], its conductivity should be not less than conventional electrolyte solutions are able to provide, *i.e.* roughly 1 S/m. Indeed, this challenge looks realistic because RTIL comprises exclusively ions. Since the conductivity of the pure RTIL is quite small, acetonitrile is expected to increase it by decreasing viscosity of the system.

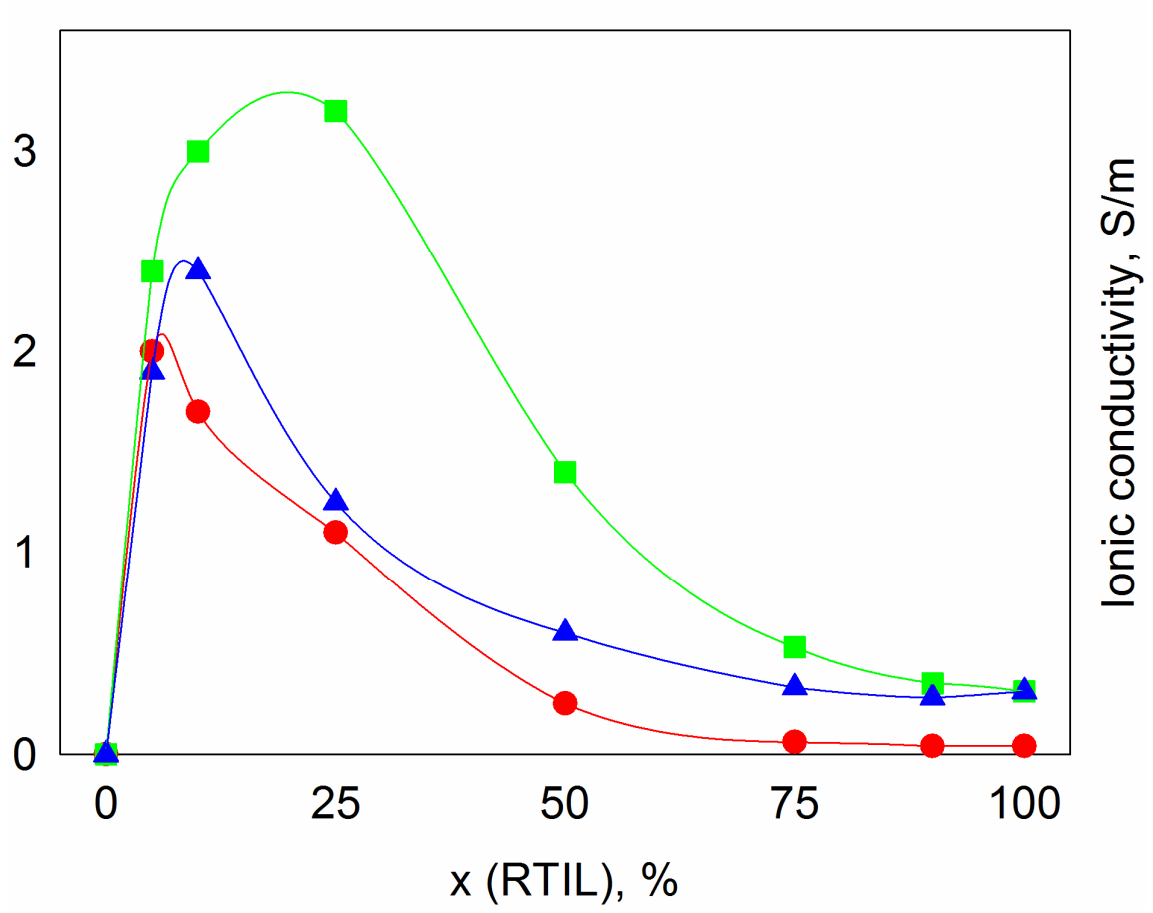

Figure 6. Ionic conductivity computed at 298 K and 1 bar for model 1 (red circles), model 2 (green squares), and model 3 (blue triangles).

Based on the current simulation (Figure 6), σ can be increased by *ca.* 10 times (from 0.31 to 3.2 S/m, model 2) provided that the molar fraction of [BMIM][$BF_4$] decreases from 100% to 25%. In conjunction with η=2.3 cP at $x$([BMIM][$BF_4$])=25%, such enhanced conductivity greatly favors the application of the mixtures of [BMIM][$BF_4$] and ACN as an electrolyte. Remarkably, the position of the peak in Figure 6 depends on the model of RTIL. As the electrostatic charges on RTIL increase, the maximum shifts to smaller molar fractions. In the case of model 1, the maximum is located at $x([BMIM][BF_4]) < 5\%$ and this significantly differs from models 2 and 3 providing a better description of the ionic dynamics in the RTIL+ACN mixtures.

The conductivity of any electrolyte solution strongly depends on the diffusion of the counterions. An ideal electrolyte can be imagined as an electrolyte comprising exclusively ions, whose motion is however independent, *i.e.* the inter-ionic interactions are marginal. The ionic liquids fulfill the first criterion, but the motion of their counterions is greatly correlated, generally resulting in a glass-like behavior of the substance. The admixtures of the aprotic solvent to some extent screen electrostatic interactions and increase ionic conductivity by several times, notwithstanding the total number of ions per volume unit decreases. In order to understand the conductivity decrease due to the inter-ionic interactions, it may be helpful to assess the conductivity using the hypothetic assumption that the ions in the RTIL-rich mixtures move independently. In the framework of the classical molecular dynamics simulations, this can be accomplished by applying the Nernst-Einstein equation. We derive the diffusion constants of the freely moving ions from the simulation of the infinitively dilute solutions of $BMIM^+$ and $BF_4^-$, containing a single ion each. Although the concentration of the ion in such a solution is not negligible, the inter-ionic interactions are absent; therefore, the ion formally exists in a solution at the infinitive dilution. The calculated diffusion constants of $BMIM^+$ and $BF_4^-$ are 1.2 and 1.5 ($\times 10^{-9}$ $m^2$ $s^{-1}$), respectively. The corresponding molar conductivity at the infinite dilution is *ca.* 100 S $cm^2$ $mole^{-1}$. In order to obtain the hypothetic conductivities for real mixtures, the diffusion constants of the ions are assumed to be equal to *D*s at infinite dilution. Figure 7 depicts real conductivities *versus* these hypothetic conductivities as a function of the system molarity. Expectedly, there is a huge difference between these sets of data indicating that inter-ionic motion plays a crucial role in the real systems. For instance, the corresponding ratios are *ca.* 4 for the 5% RTIL mixture and *ca.* 170 for 90% RTIL mixture. Note that even 5% RTIL mixture is not a dilute solution, since the electrolyte concentration in it is almost 0.9 mole $dm^3$, while the own concentration (100% RTIL) of $[BMIM][BF_4]$ is 5.9 mole $dm^3$. The ionic conductivity in the real electrolyte systems exhibits quite a complicate dependence on the concentration due to the non-ideal behavior. In this context, the atomistic simulation is often the only available theoretical algorithm to derive this function with a quantative quality.

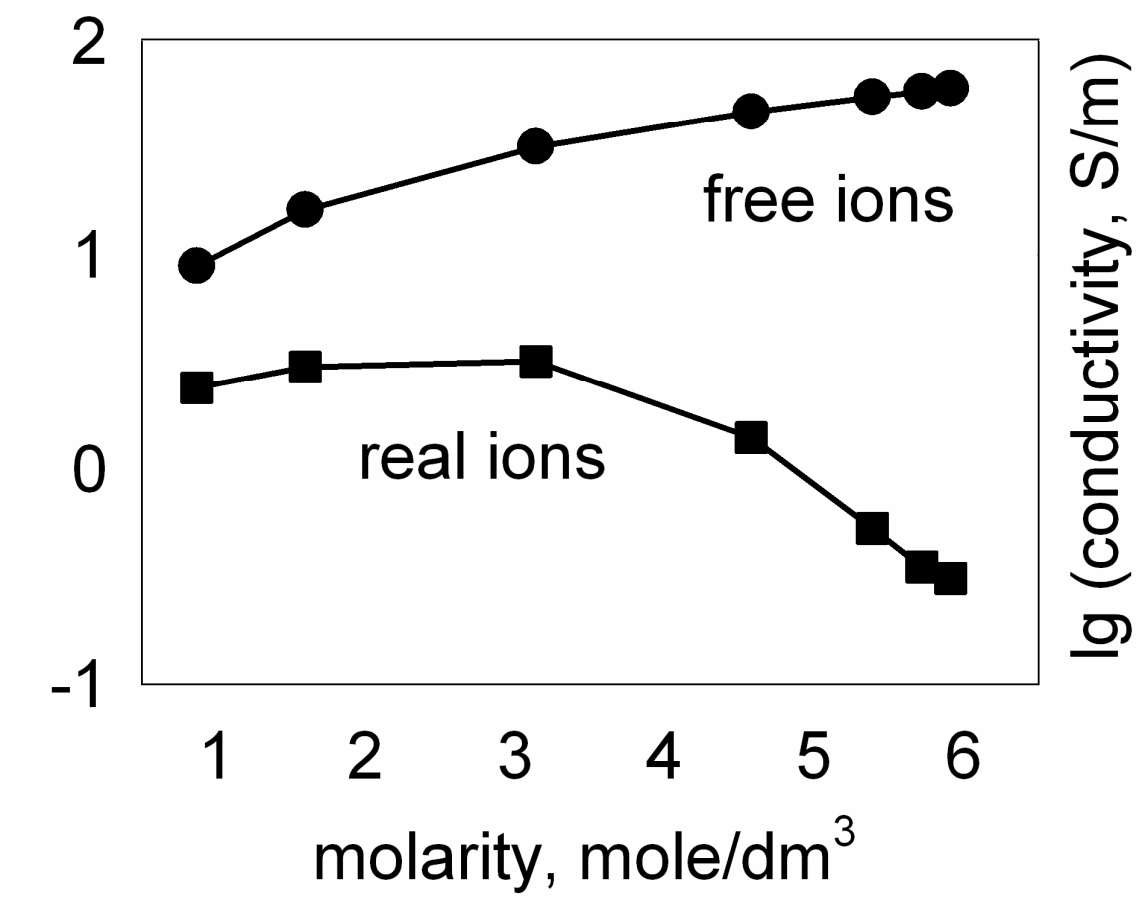

Figure 7. The logarithm of the ionic conductivity computed using the ionic diffusion of freely moving ions (circles) and real ionic conductivity derived from the Einstein-Helfand method (squares) using molecular dynamics simulations (model 2).

The heat of vaporization, $H_{vap}$, (Figure 8) directly reflects ionic, molecular and ion-molecular interaction energies in the tightly bound system. Unfortunately, the experimental data of $H_{vap}$ for [BMIM][$BF_4$] + ACN systems is not yet available. Although both of them are liquids with dominating Coulombic (strong) interactions, $H_{vap}$ of [BMIM][$BF_4$] is *ca.* four times larger than of ACN. This results in a strong linear dependence of $H_{vap}$ of RTIL-ACN upon the mixture composition. The slopes are 1.5, 1.1, and 1.1 kJ/mol for models 1, 2, and 3, respectively, whereas all of the correlation coefficients exceed 99%. Again, the observed difference among the original model 1 and its derivatives, proposed in this work, is quite significant. On the other hand, the numeric difference between $H_{vap}$s from different models is moderate notwithstanding a significantly reduced non-bonded energy. This implies that the systems simulated with model 2 tend to expand in order to increase their entropy.

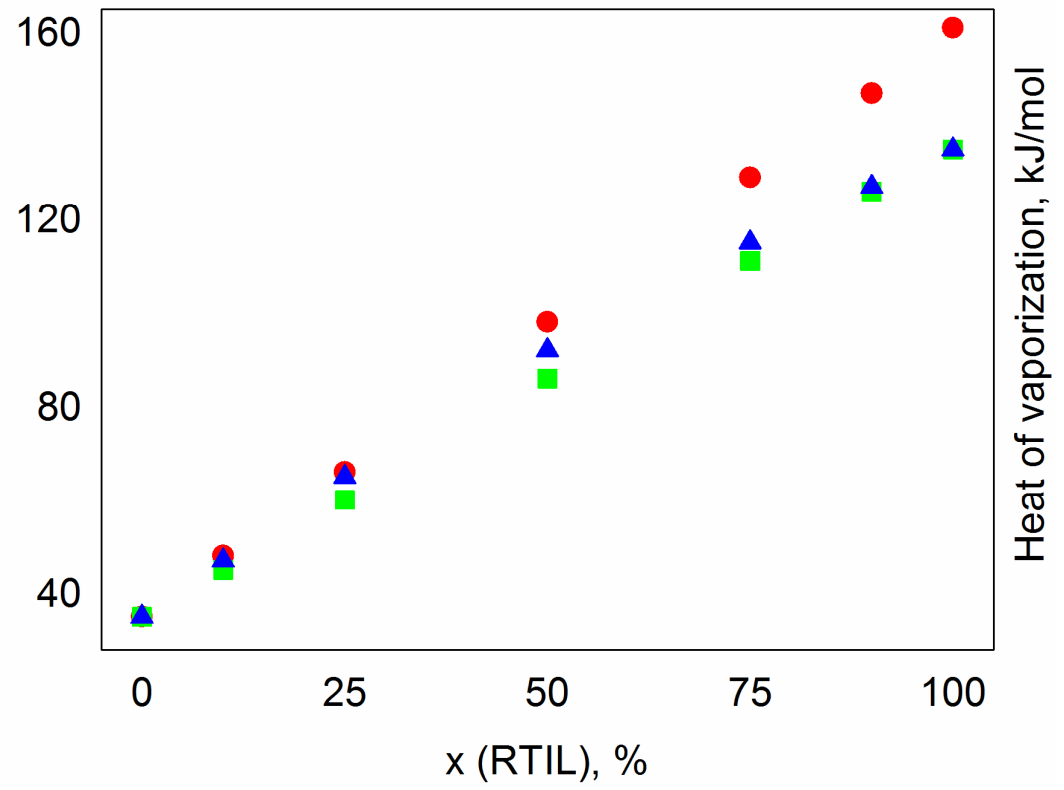

Figure 8. Heat of vaporization of the [BMIM][$BF_4$] + ACN mixtures computed at 298 K and 1 bar for model 1 (red circles), model 2 (green squares), model 3 (blue triangles).

The radial distribution functions (RDFs) between the hydrogen atoms of the imidazole ring (often referred as HA) of the cation and the fluorine atoms of the anion, $g_{HF}(r)$, (Figure 9) reflect an ionic structure in the mixtures of varying compositions. The well-defined first maximum at 0.22 nm predicts a preferencial existence of the contact ionic pairs, rather than acetonitrile-separated ions, over the whole composition range. Interestingly, as the molar fraction of RTIL increases from 10 to 90%, the height of the first maximum decreases from 4.2 to 2.5 (according to model 2). Owing to mixing, the molecules of ACN substitute ions in the ionic clusters, and consequently, enforce cation-anion interactions since the interactions between ACN and ions are substantially weaker than between counterions. Generally, this results in a higher degree of the correlations of ionic motion. Such a finding is well underlined by the dependence of cationic and anionic diffusion coefficients on the molar fraction of [BMIM][$BF_4$]. Indeed, at $x$(RTIL) > 25%, $D_+$ and $D_-$ are almost indistinguishable and small. In turn, it preserves a maximum of ionic conductivity at $x$ (RTIL) < 25%. The second maximum occurs at 0.41 nm which coincides well with the second maximum on $g_{HF}(r)$ for the pure [BMIM][$BF_4$][27] suggesting the existence of larger ionic clusters in the mixtures. This observation even includes the mixture with $x$=10% of RTIL. Similarly to the first maximum, the second maximum decreases from 2.0 to 1.4 as the molar fraction of [BMIM][$BF_4$] increases from 10 to 90%. To recapitulate, mixing of [BMIM][$BF_4$] with ACN enforces stability of the ionic pairs, and this effect is in inverse proportion to the content of RTIL in the mixture. Our findings are also supported by the recent studies of Wu *et al.*[49] who used a somewhat different model for ACN.

In addition, the structure of RTIL and acetonitrile can be characterized by a radial distribution between the nitrogen atom of ACN and the hydrogen (HA) atom of the imidazole ring of [BMIM]$^+$, $g_{HN}(r)$. The positively charged HA attracts the most negative interaction site of ACN. The resulting RDF can be used to study the formation of solvation shells of the cation as a function of the ACN content. As suggested by Figure 10, the inter-ionic interactions are more preferrable than ion-molecular interactions over the whole composition range in the $[BMIM][BF_4]$+ACN mixture. Indeed, both maxima on $g_{HN}(r)$, though pronounced, are appreciably smaller than on the cation-anion RDF, $g_{HF}(r)$ (1.8 *vs*. 4.2 and 1.2 *vs*. 2.0 for the 10% RTIL mixture, model 2). Unlike $g_{HF}(r)$, the peak values of $g_{NF}(r)$ are almost insensitive to the composition of the mixture. The difference among inter-ionic ($BMIM^+ – BF_4^-$) and ion-molecular ($BMIM^+$ – ACN and $BF_4^-$ – ACN) interaction energies per mole prevents ACN from a complete dissolving of $[BMIM][BF_4]$. Both $g_{HF}(r)$ and $g_{HN}(r)$ indicate that there are no solvent-separated ion pairs in the mixtures. Meanwhile, the first and the second solvation shells of $[BMIM]^+$ are stable.

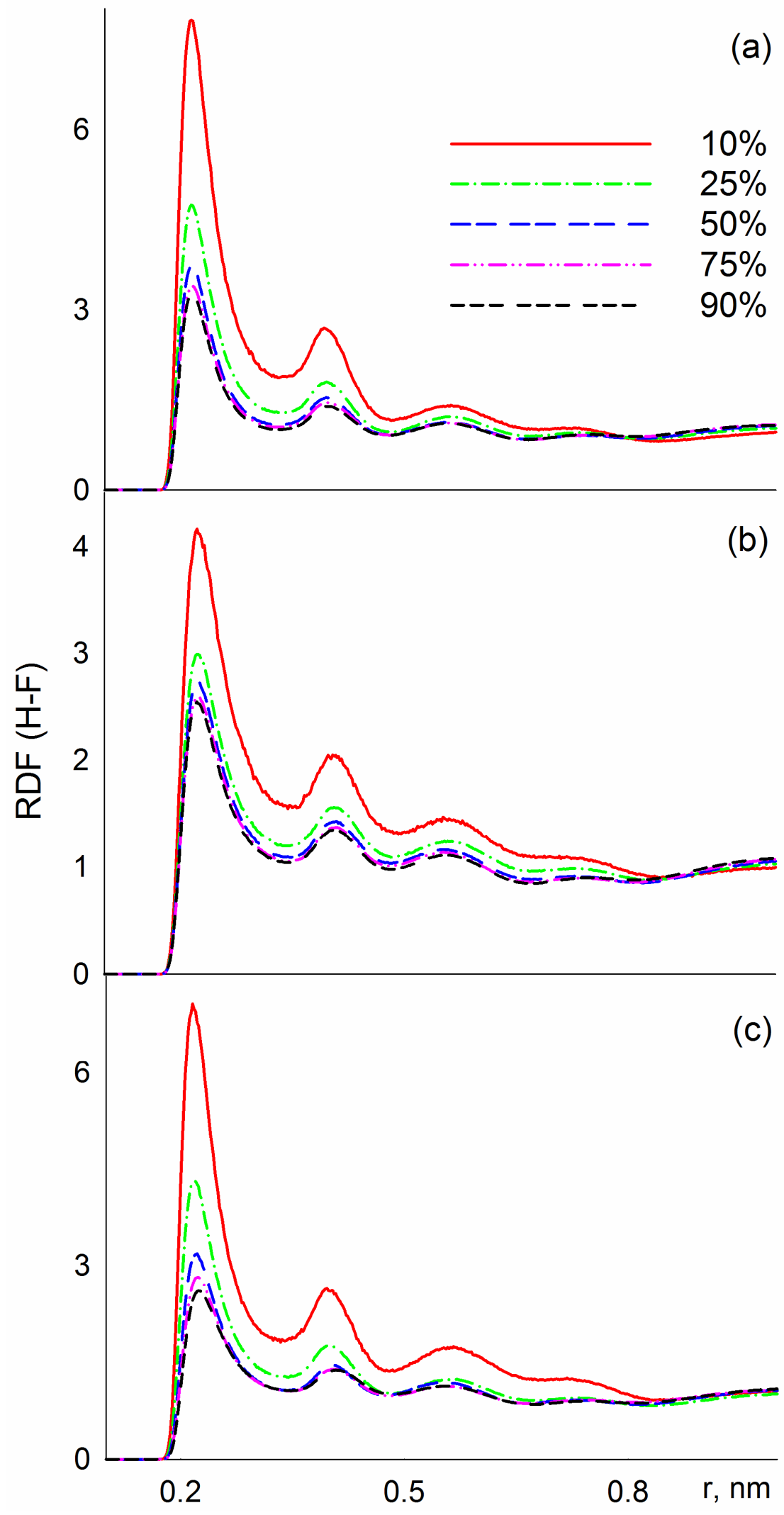

Figure 9. Radial distribution functions of the hydrogen (HA) site of $BMIM^+$ and fluorine (F) site of $BF_4^-$ for model 1 (a), model 2 (b), and model 3 (c). The legend shows the molar fractions of RTIL in the mixtures.

While all three models agree in qualitative predictions of the structure properties, the values of $g_{HF}(r)$ and $g_{HN}(r)$ vary. For instance, the decrease of $f_{sc}$ down to 0.81 leads to a weaker peak on $g_{HF}(r)$ (4.1 *vs*. 7.8). This is expected since the interaction energy among ions is affected very significantly by scaling. In the case of $g_{HN}(r)$, the effect is not evidenced so much and is comparable with the impact of the molar fraction of RTIL. One may conclude that the structure and stability of the solvation shells of $BMIM^+$ is, to some extent, not sensitive to its electrostatics. Overall, the impact of the scaling factor is about an order of magnitude larger for transport properties than for structure and thermodynamics. Such a non-uniformity creates a background to tune force field models of the RTIL-containing systems by adjusting only Coulombic contribution to the total potential energy.

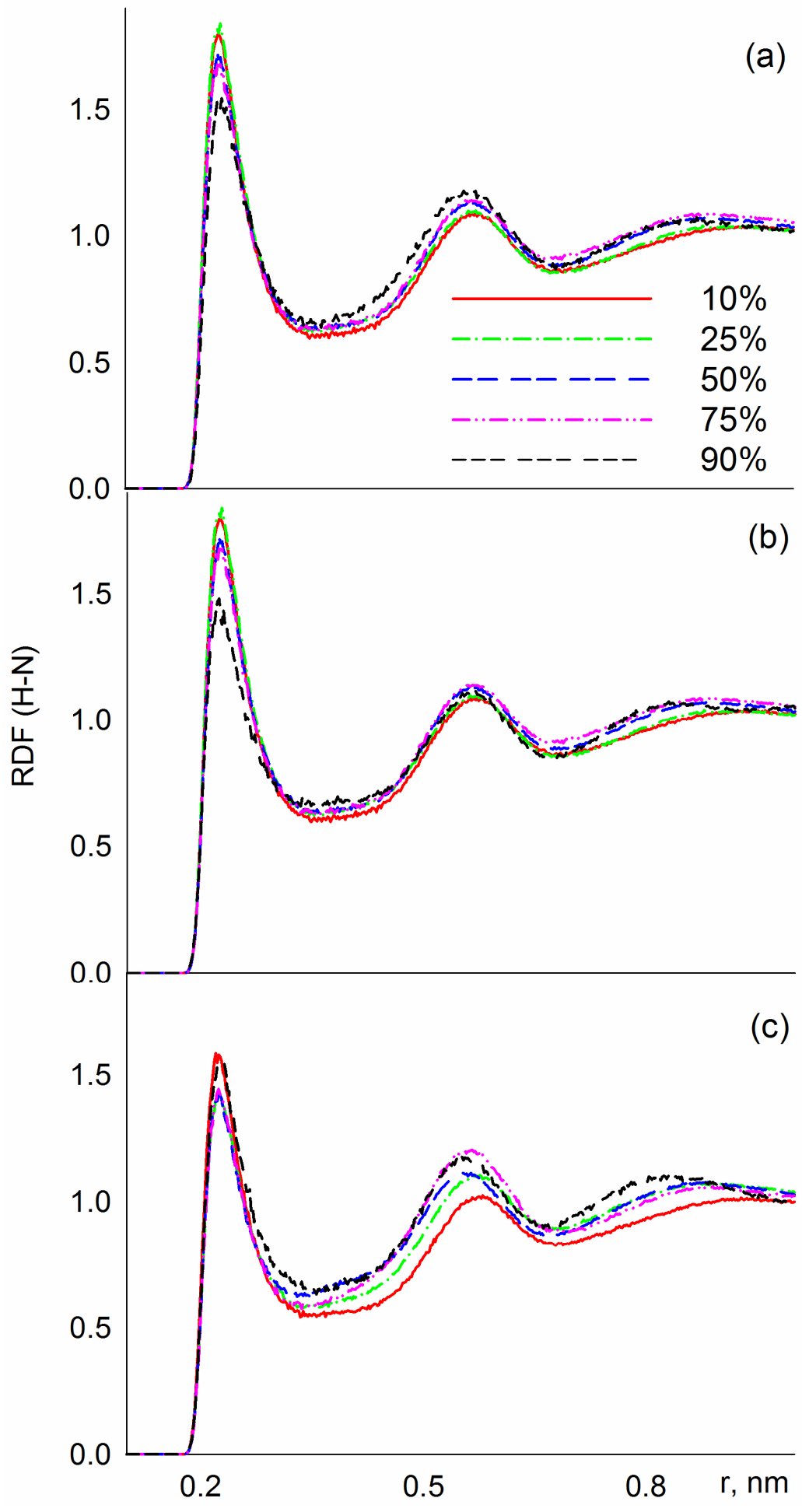

Figure 10. Radial distribution functions of the hydrogen (HA) site of $BMIM^+$ and nitrogen (N) site of ACN for model 1 (a), model 2 (b), and model 3 (c). The legend shows the molar fractions of RTIL in the mixtures.

## Conclusion

A series of classical molecular dynamics simulations of the mixtures, containing 1-butyl-3-methylimidazolium tetrafluoroborate and acetonitrile, are carried out. Three approaches to constructing the force field model of this mixture are considered. Whereas all three treat van der Waals interactions and the thermal oscillations of bonds and angles identically, the electrostatic interactions are introduced differently. The first model exploits the electrostatic potential around ions derived from the high level *ab initio* calculations in vacuum, *i.e.* with unit charges on the cation and the anion. The second model uses the recent findings that electrostatic charges on all ions should be equally reduced to phenomenologically account for an electronic polarization in condensed phase. Herewith, we assume that the polarization of the cation and the anion estimated in the system of pure RTIL can be directly transferred to the mixtures of this RTIL with an aprotic solvent. In other words, it is supposed that the ions are evenly polarized by each other as well as by neutral molecules (acetonitrile). On the contrary, in the third model we assume that acetonitrile does not polarize RTIL at all. Therefore, in the third model the mixture should be simulated by means of the floating-charge model where the partial charges on each interacting site functionally depend on the molar fraction of the ionic liquid.

The simulation of thermodynamic, structure and transport properties shows that both models proposed in this work, *i.e.* models 2 and 3, perform reasonably well over the whole range of compositions. While the standard procedure leading to model 1 is acceptable for reproducing properties such as density and pair distribution functions, it is unable to reproduce ionic transport. At $x$ ([BMIM][$BF_4$]) < 25 %, model 3 systematically overestimates shear viscosity, while model 2 remains successful. This suggests that ions are polarized not only by each other but also to a large degree by the acetonitrile molecules. The maximal ionic conductivity of the [BMIM][$BF_4$]+ACN is found at 10 % < $x$ ([BMIM][$BF_4$]) < 25 %. Its dependence on molar fraction agrees well with the other properties of these mixtures. In turn, the decrease of the molar fraction of [BMIM][$BF_4$] enforces the stability of the ionic aggregates, including ion pairs and larger structures.

**Acknowledgements**

Vitaly V. Chaban is grateful to Amanda J. Neukirch (University of Rochester) for a number of valuable considerations. The research is supported in part by the NSF grant CHE-1050405.